\documentclass[useAMS,usenatbib,referee]
{biom_arxiv}

\usepackage{bbm}
\usepackage{color}
\usepackage{amsmath}
\usepackage{graphicx}
\usepackage{caption}
\usepackage{subcaption}
\usepackage{url}
\usepackage{xcolor}
\usepackage{mathtools}
\usepackage{enumitem} 
\usepackage{setspace}
\usepackage{amsmath, amssymb}
\usepackage{booktabs}
\usepackage{tabularx}
\usepackage{array}
\newcommand{\estp}[2]{%
  \begin{tabular}[c]{@{}c@{}}#1\\[-2pt]{\color{gray}\scriptsize #2}\end{tabular}%
}
\usepackage{xr}

\definecolor{color1}{HTML}{FF7F50}
\definecolor{color2}{HTML}{6610F2}
\definecolor{color3}{HTML}{4CBB17}

\def\bSig\mathbf{\Sigma}

\title{Beyond the Composite: Enhancing Trial Analysis through a\\ Divide \& Conquer Approach to `Days Alive and at Home':\\ Insights from the NOTACS trial}

\author{
{\normalfont\small Preprint version: 22 May 2026}\\
\smallskip
Letao Yuan$^{1,*}$\email{letao.yuan@mrc-bsu.cam.ac.uk}, 
Sofía S. Villar$^{1,2}$, and
Dominique-Laurent Couturier$^{1,2}$ \\
$^{1}$MRC Biostatistics Unit, University of Cambridge, Robinson Way, Cambridge, CB2 0SR, U.K. \\
$^{2}$Papworth Trials Unit Collaboration, Royal Papworth Hospital, Cambridge, U.K.}

\begin{document}
\pagestyle{plain}



\pagerange{\pageref{firstpage}--\pageref{lastpage}} 
\pubyear{2026}
\artmonth{May}


\label{firstpage}

\begin{abstract}
``Days alive and at home" (DAH) is a recent patient-centered outcome measure for perioperative trials, defined as the number of days a patient spends at home during the follow-up period. DAH typically follows a zero-inflated, left-skewed, bi-modal distribution. Other increasingly used complex endpoints, such as days alive without a ventilator, share these statistical features arising from combining survival with another clinically relevant count outcome into a single, comprehensive measure. A key challenge for DAH and similar endpoints is the lack of a readily identifiable distributional form, which complicates the statistical design of trials using it as the primary endpoint, particularly regarding the robustness of sample size calculations and final analyses where the central limit theorem might not be suitable. Using 200 data points from the interim data of the NOTACS trial (ISRCTN14092678), whose primary endpoint was DAH, we developed a novel `Divide \& Conquer' model that breaks DAH into distinct parts modeled individually. To our knowledge, such a model has not been used before for DAH. We demonstrate that our approach significantly improves model fit compared to existing alternatives, enabling more suitable DAH data generation that can be used for simulation-based sample size calculations and evaluation of operating characteristics of the statistical test(s). Beyond NOTACS, our work has large potential to inform the design and analysis of other trials using DAH or similar complex endpoints.

\end{abstract}

%



\maketitle
\thispagestyle{plain}
\pagestyle{plain}

%

\section{Introduction}
\label{sec:intro}

``Days alive and at home" (DAH) after a hospital stay is a recently-developed, patient-centered outcome measure. It was initially utilized to estimate outcomes in clinical trials for chronic \citep{ariti2011} and acute diseases \citep{fanaroff2018}, and has recently been introduced and validated as an endpoint in perioperative trials \citep{myles2017}. 

DAH is the total number of days a patient spends at home during the post-intervention follow-up. The follow-up duration varies by intervention - often 30 or 90 days for cardiac surgery - and is typically appended to the acronym (e.g., ``DAH90'' for a 90-day period).

Previously, researchers calculated DAH by subtracting both hospital days and the days following a patient's death from the total follow-up period. \cite{myles2017} modified the definition of this outcome measure by setting the DAH to 0 for any patient who died within the follow-up period, effectively giving death the same weight as being hospitalized or away from home for the entire duration. Furthermore, the days spent in nursing facilities were also excluded from the calculation, restricting the endpoint strictly to time spent at home. This refined definition, which assigns a zero score to death and recognizes that time in a nursing facility differs clinically from being at home, has become widely adopted in recent surgical trials that use DAH as an endpoint. The NOTACS trial that motivated this work also relied on Myles's definition of DAH, as detailed in Section \ref{sec:notacs}. 

Compared to the Length of Stay (LOS) endpoint, DAH not only measures delayed hospital discharges, but also integrates clinically important post-discharge information, such as readmission, mortality, and transfers to nursing facilities due to postoperative complications into a single patient-centered outcome metric. DAH helps avoid potentially misleading conclusions associated with LOS. As noted by \cite{bueno2010} and \cite{carey2014}, early discharges do not always indicate fast recovery and can be linked to higher readmission or mortality rates. A trial relying solely on LOS might inappropriately favor a treatment that shortens initial stays but increases total hospital time or readmissions. By capturing these nuances, DAH provides a more patient-centered assessment of treatment effects.

Importantly, in certain settings, such as surgical trials, blinding of clinical teams and patients is impossible due to the nature of the intervention, thus making it possible for clinicians to consciously or unconsciously discharge patients receiving their preferred treatment earlier, while allowing patients to alter their behavior and recovery expectations based on their known allocation. Through its treatment of death, readmissions and stays in nursing facilities, DAH can help mitigate such potential biases.

Moreover, DAH is more robust to small, clinically insignificant variations in discharge timing. For instance, administrative delays may postpone discharge by a day or two even when a patient is clinically ready. While such delays can disproportionately skew LOS, especially in trials where average stays are short, their impact on DAH is diluted within the context of the total follow-up period. Consequently, DAH provides a more stable reflection of the overall treatment effect by reducing the ``noise'' of minor administrative variances.

The use of DAH has markedly risen recently. Due to the way to define it - based on information related to death, LOS, readmission and nursing care - DAH follows a complex distribution, characterized by
\begin{itemize}[itemsep=-5pt,topsep=0pt, leftmargin=1cm]
    \item a minimum post-intervention hospital length of stay, which is expected to vary depending on the type of intervention and to be hospital-specific, 
    \item a clump-at-zero, due both to patients who died and to patients who did not spend a day at home during the follow-up period (as a consequence of long hospital stays or discharges to nursing homes),
    \item left-skewness, as most patients are typically discharged shortly after surgery while others require extended hospital care, 
    \item a potential second mode of strictly positive values possibly due to readmissions. 
\end{itemize}

The NOTACS interim dataset, described in Section \ref{sec:notacs}, illustrates these features.  Similar characteristics can also be observed in other studies, for example, in \cite{reilly2022}, where a second mode of positive values can clearly be identified for interventions such as hip and knee joint replacements (Figure 5 of their supplementary material).

This inherent complexity of DAH often leads to poor model fits under standard approaches, a problem we illustrate in Section \ref{sec:model}. This lack of fit is not merely a theoretical issue; it can result in incorrect power projections during the trial design phase. 

In this work, we therefore introduce a novel parametric modeling approach for DAH that can be used for other complex endpoints. By providing a more accurate reflection of the data-generating process, our model offers a more reliable framework for the design and analysis of such trials. The paper is structured as follows: Section \ref{sec:notacs} introduces the NOTACS trial, the motivating data source for this work, and describes the key distributional features of DAH. Section \ref{sec:model} describes existing models for DAH and introduces our novel model. Section \ref{s:notacs-application} compares the goodness-of-fit of different models based on the interim data from NOTACS. Section \ref{sec:modelapplication} describes how our model can be used, for example, to evaluate the operating characteristics of statistical tests considering DAH and perform sample size calculations. Finally, strengths and limitations of our work as well as future research directions are discussed in Section \ref{s:discuss}.

\section{Data Source: the NOTACS Trial}
\label{sec:notacs}

This project used the data from 200 randomly selected patients in the NOTACS trial's interim analysis to develop improved modeling for DAH.

The NOTACS (Nasal High-Flow Oxygen Therapy After Cardiac Surgery) trial is an adaptive, international multicenter, parallel-group randomized controlled trial (RCT), designed to investigate the efficacy, safety and cost-effectiveness of prophylactic use of high-flow nasal therapy (HFNT) for patients undergoing cardiac surgery with cardiopulmonary bypass. Detailed information on trial design can be found in the published study protocol and statistical analysis plan for the NOTACS trial \citep{earwaker2022,dawson2022}. 

DAH90 was the primary endpoint in the NOTACS trial, defined as the number of days spent at home within a 90-day follow-up period. Following \cite{myles2017}, DAH was recorded as zero for any patient who died during this window. Notably, days spent in nursing care were considered as ‘home’ only for patients who had resided there prior to surgery.

\begin{figure}[!htbp]
    \includegraphics[width=0.9\textwidth, height=0.9\textheight, keepaspectratio]{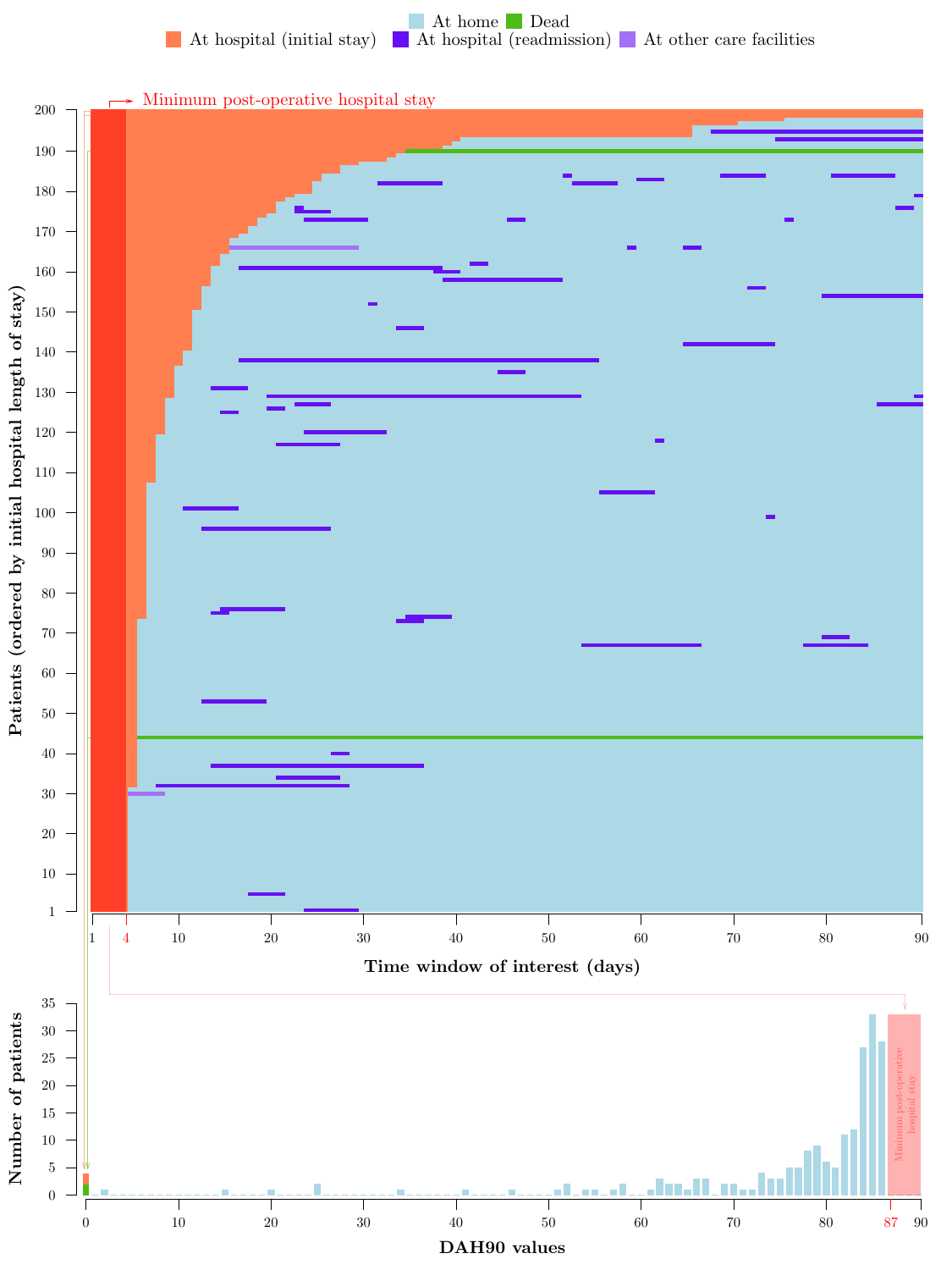}
    \centering
    \caption{Upper panel: Color-coded location of each patient (y-axis) for each day of the 90-day period following intervention (x-axis). Patients are ordered by their postoperative length of hospital stay. Lower panel: Barplot of DAH90. The minimum postoperative stay (red) restricts maximum DAH90 values to 86. The clump-at-zero is attributed to both patient mortality (green, patients 44 and 190 in the upper panel) and prolonged initial stays surpassing the 90-day window (orange, patients 199 and 200 in the upper panel).}
    \label{fig:fig1}
\end{figure}

The upper plot of Figure \ref{fig:fig1} shows the color-coded location of each patient (y-axis) for each day of the 90-day period following intervention (x-axis). Patients are ordered by their initial length of hospital stay. We can note that, for the NOTACS intervention, (i) there is a minimal length of hospital stay of 4, (ii) patients 44 and 190 died (green) and patients 199 and 200 have a post-intervention length of stay exceeding 90 days leading to DAH values of zero for 4 patients, (iii) more than half of the patients (patients 1 to 119) left the hospital within a week after the intervention and the hospital discharge of the remaining patients is skewed, ranging from 8 to 90 days, at which point data are right-censored, (iv) readmission or other care occurred for almost a quarter of  patients (n=47). 
The lower panel of Figure \ref{fig:fig1} shows the corresponding barplot of DAH90, in which the minimum postoperative stay restricts maximum DAH90 values to 86. We can note the presence of a clump-at-zero induced by mortality and post-intervention lengths of stay exceeding 90 days, left-skewness, and a possible second mode related to post-discharge subsequent care.

\clearpage

\begin{figure}[!htbp]
\centering
\includegraphics[width=0.65\textwidth, keepaspectratio]{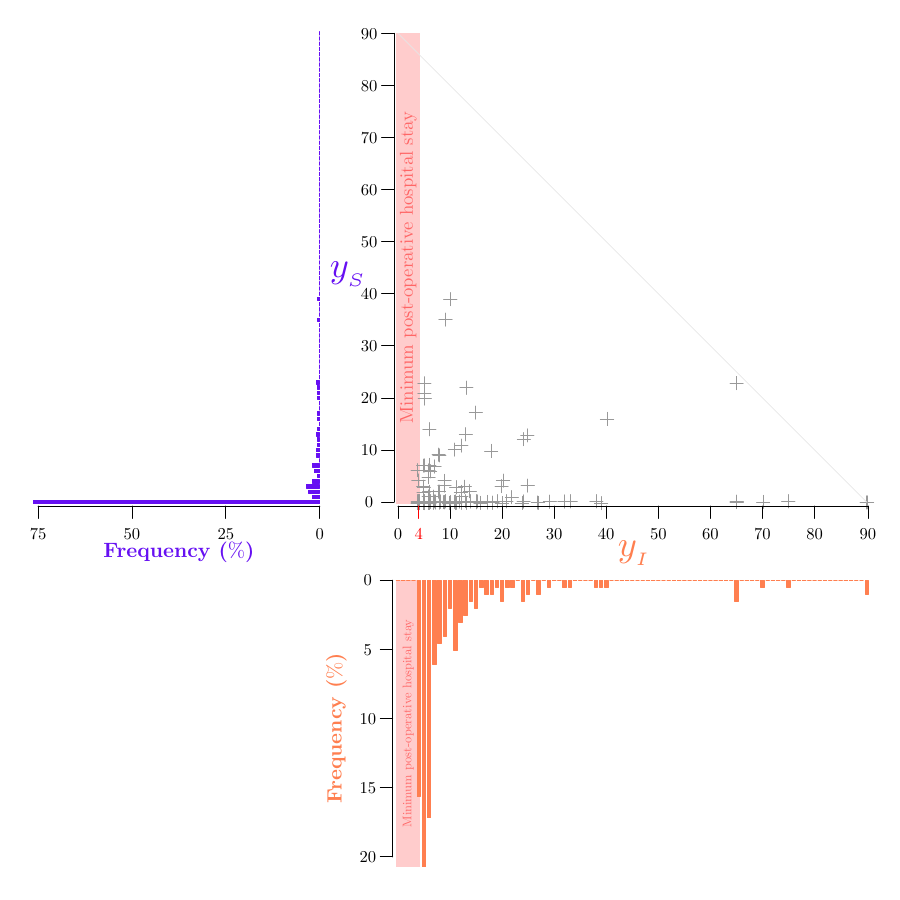}
\caption{Barplots of the postoperative initial stay, $y_{_{I}}$ (lower right), and post-discharge subsequent care, $y_{_{S}}$ (upper left), and scatterplot of the relationship between these two random variables (upper right) for the 198 patients of the NOTACS interim analysis that were alive at the end of the study. The oblique gray line in the scatterplot shows the limit above which no point can't be observed because of the constraint $y_{_{I}}+y_{_{S}}< u$.}
\label{fig:fig2}
\end{figure}

The lower and upper left panels of Figure \ref{fig:fig2} respectively display the barplots of the postoperative initial hospital stay (orange, denoted $y_{_{I}}$) and of the post-discharge subsequent care (blue, denoted $y_{_{S}}$). The postoperative length of stay has a lower bound of 4 days, reflecting the minimum hospital stay, and an upper bound of 90 days. Its distribution is right-skewed, with few occurrences (fewer than 10\%) exceeding 25 days. The barplot for post-discharge subsequent care exhibits a clump-at-zero of approximately 75\%, corresponding to patients who did not require care following the first hospital discharge, and a skewed distribution of strictly positive values, ranging from 1 to 40 days. The upper right panel displays the scatterplot of the relationship between these two random variables, which is constrained by the boundary $y_{_{I}}+y_{_{S}}< u$ (indicated by the oblique gray line).

\clearpage

\section{Methodological Framework: Statistical Models for DAH}
\label{sec:model}
\subsection{Review of In-use DAH Modeling Strategies }

From a modeling perspective, DAH presents four challenges: (i) left skewness, (ii) zero inflation, (iii) bimodality, and (iv) a minimum postoperative hospital stay:

\textit{(i) Left-skewness:} Many studies address the left-skewness of DAH by instead modeling its right-skewed complement, the total days spent away from home. In the absence of readmissions or intermediate care, this variable is equivalent to the initial hospital length of stay (LOS). Examples are the log-normal, considered when designing NOTACS trial \citep{earwaker2022,myles2018}, and count distributions like the Poisson \citep{suikkanen2021}, Negative Binomial \citep{vanhoutven2019}. Other approaches include the ones of \cite{wu2022}, who used a Beta-Binomial distribution to model the number of days alive and at home during the postoperative follow-up window, and \cite{goldberg2013}, who considered a Beta regression to model the proportion of time spent at home during the follow-up period. 

\textit{(ii) Zero-inflation:} DAH is characterized by two sources of zeroes, the ones induced by death and the ones induced by LOS exceeding the follow-up period. The zero-inflation in DAH is typically addressed using models suited for binary data, such as logistic regression, to account for the probability of a zero outcome.

\textit{(iii) Bimodality:} A less appreciated characteristic of DAH lies in its distribution of strictly positive values. Depending on the context, DAH can show bimodality, for example due to readmission stays. To the best of our knowledge, all existing approaches consider a single statistical distribution to model strictly positive DAH values. This might not allow to adequately capture the characteristics of a distribution induced by distinct underlying processes, such as the length of postoperative hospital stay, the probability of readmission and the length of readmission time.

\textit{(iv) Minimum hospital stay:} 
In clinical practice, a minimum postoperative stay is typically required after an invasive surgery to allow close monitoring of patient recovery. This duration depends on the intervention and the discharge protocol, and differs from prolonged stays due to intervention-related complications or poorer health status. This implies that the minimum postoperative hospital stay cannot be zero. However, to the best of our knowledge, this feature has not been explicitly accounted for in existing approaches. An exception to this is \cite{myles2021} who seems to be using an inferred protocol time.

Table \ref{tab:existingmodel} summarizes existing distribution-based models for DAH.

\begin{table}[htbp]
\centering
\singlespacing
\small
\caption{Existing modeling approaches for DAH}
\label{tab:existingmodel}
\setlength{\tabcolsep}{5pt}
\renewcommand{\arraystretch}{1.25}

\begin{tabularx}{\textwidth}{@{}
    >{\raggedright\arraybackslash}p{3.5cm}
    >{\raggedright\arraybackslash}p{6cm}
    >{\raggedright\arraybackslash}X
@{}}
\toprule
\textbf{Reference} &
\textbf{Methodological Approach} &
\textbf{Stochastic components} \\
\midrule

\multicolumn{3}{@{}l}{\textit{Zero-adjusted direct models: Combine a logistic component for zeros with a direct model for positive DAH values.}} \\
\addlinespace[2pt]

Wu 2022
& Zero-adjusted beta-binomial model
& $\begin{array}[t]{@{}l@{}}
\mathbbm{1}(R^{\dagger}_i=1) \sim \mathrm{Bernoulli}(\pi_{R^{\dagger}_i}),\\
Y_i^u \mid (R^{\dagger}_i=1) \sim \mathrm{BBtr}(\mu_i,\sigma_i\mid n=u).
\end{array}$ \\

\addlinespace[4pt]

Goldberg 2013
& Zero-adjusted beta model
& $\begin{array}[t]{@{}l@{}}
\mathbbm{1}(R_i=1) \sim \mathrm{Bernoulli}(\pi_{R_i}),\\
\frac{Y_i^u}{u} \mid (R_i=1) \sim \mathrm{Beta}(\mu_i, \phi_i).
\end{array}$ \\

\midrule

\multicolumn{3}{@{}l}{\textit{Flipped models: Model non-home days before transforming back to DAH; may incorporate a logistic zero-component.}} \\
\addlinespace[2pt]

Myles 2018; Litton 2026
& (Zero-inflated) (flipped) log-normal model
& $\begin{array}[t]{@{}l@{}}
Y_i^u = (u-Y_{H_i})\,\mathbbm{1}(D_i=0),\\
Y_{H_i} \sim \mathrm{LogNormal}(\mu_i,\sigma_i),\\
\mathbbm{1}(D_i=1) \sim \mathrm{Bernoulli}(\pi_{D_i}).
\end{array}$ \\

\addlinespace[4pt]

Suikkanen 2021
& Zero-inflated (flipped) Poisson model
& $\begin{array}[t]{@{}l@{}}
Y_i^u = u-Y_{H_i},\\
Y_{H_i} \sim \mathrm{PO}(\mu_i).
\end{array}$ \\

\addlinespace[4pt]

Van Houtven 2019
& (Flipped) negative binomial model
& $\begin{array}[t]{@{}l@{}}
Y_i^u = Y_{A_i}-Y_{H_i},\\
Y_{H_i} \sim \mathrm{NBI}\{\mu_i(\mathrm{offset}_i),k\}.
\end{array}$ \\

\bottomrule
\end{tabularx}

\vspace{0.5em}
\begin{minipage}{\textwidth}
\footnotesize
\textit{Notes.}
$Y_i^u$ is DAH within $u$ days for patient $i$; $Y_{H_i}$ and $Y_{A_i}$ are the corresponding hospital days and days alive, each bounded by $[0,u]$.
$\mathbbm{1}(\cdot)$ denotes the indicator function. $R^{\text{\tiny$\dagger$}}_i$, $R_i$, and $D_i$ respectively denote the binary variables `alive and returned home', `returned home', and `dead during follow-up', with corresponding probabilities $\pi_{{\text{\tiny R}_i^{\text{\tiny$\dagger$}}}}$, $\pi_{R_i}$ or $\pi_{D_i}$.
The parameters $\mu_i$ and $\sigma_i$ denote model-specific location/mean and scale/dispersion terms; $\phi_i$ is beta precision, $k$ is the negative-binomial shape parameter, $\mathrm{offset}_i$ is an the offset term accounting for maximum possible home days prior to censoring, and $n=u$ is the beta-binomial denominator.
BBtr, PO, and NBI denote the truncated beta-binomial, Poisson, and negative-binomial distributions. Parentheses in model names indicate features inferred from the published description rather than stated explicitly.\\
Parentheses in the second column indicate we could not find explicitly in the original publication whether excess zeros were modeled separately, or whether the assumed distribution fitted days not at home or other components in DAH. We made informed assumptions in these cases. 
\end{minipage}
\end{table}

Some studies have considered quantile \citep{myles2017,myles2018} and ordinal \citep{shinall2023, waddingham2025} regressions to analyze DAH. While these methods are likely to achieve the desired type I error control and to be suitable in specific cases, we chose not to investigate them here, as they do not, \emph{per se},  allow suitable modeling in all cases we are interested in. Indeed, these approaches assume that zero and positive DAH values can be modeled via a unique linear predictor despite corresponding to different processes, like death, and short or long hospital stays. The proportional odds assumption, which requires constant covariate effects across all outcome thresholds, is unlikely to hold for DAH. As a composite of distinct clinical processes like mortality, initial hospitalization, and readmission, a single factor like age may strongly influence the probability of survival (zeros) while having a negligible impact on incremental stay durations, such as 3 versus 4 days. Also, quantile regression can show numerical instability when fitted on zero-inflated or (non-continuous) count data \citep{ling2022, alampi2025}. We also chose not to consider the approach consisting of dichotomizing DAH and modeling the resulting random variable via a method suitable for binary endpoints \citep{rasmussen2023} due to the corresponding lack of information and power.

In the following sections, we apply the  `Divide \& Conquer' strategy to develop a multi-component parametric model. We then evaluate its performance against existing approaches by comparing their respective fits to the NOTACS interim dataset.

\subsection{The ‘Divide \& Conquer’ General Modeling Framework}

In this Section, we describe how we used the `Divide \& Conquer' strategy from computer science to break DAH's complexity into several simpler parts that we modeled separately, thus allowing us to better capture the actual clinical processes generating the data.

For patients alive at the end of follow-up, we isolated the underlying clinical processes by deconstructing the total hospital length of stay into distinct components. First, we separated the initial postoperative stay from subsequent days of extra care, across hospital and other care settings, depending on the definition used. Second, we divided the initial stay itself into a structural minimum stay — dictated by hospital protocols — and a variable extension period driven by surgical complications or poor patient health. We finally handled death occurring at anytime during the follow-up period separately. 

These different components form the critical sub-parts we focused on in developing our model. Hence, our suggested DAH model is given by

\begin{equation}
y_i^u = \bigg[u-{\color{color1}\underbrace{P_i^{{\mathbbm{1}(E_i = 0)}}\left(\tilde{p}_i + y_{_{E_i}}\right)^{\mathbbm{1}(E_i = 1)}}_{y_{_{I_i}}}} - {\color{color2}\underbrace{y_{_{C_i}}\mathbbm{1}(C_i = 1)}_{y_{_{S_i}}}}\bigg]{\color{color3}\mathbbm{1}(D_i = 0)}
\label{eq:mainequation}
\end{equation}

where
\begin{itemize}[itemsep=-5pt]
\item $y_i^u$ denotes the DAH outcome for patient $i$, where $y_i\in[0,u]$ and $u$ denotes the duration of the follow-up time of interest, typically measured in days,
\item $\mathbbm{1}(\cdot)$ is an indicator function equals to 1 if the condition is satisfied and 0 otherwise,
\item {\color{color1}$y_{_{I_i}}=P_i^{{\mathbbm{1}(E_i = 0)}}\big(\tilde{p} + y_{_{E_i}}\big)^{\mathbbm{1}(E_i = 1)}$} denotes the total number of days patient $i$ spent in hospital after surgery and before the first discharge (i.e. the initial length of hospital stay), where
    \begin{itemize}[itemsep=-5pt,topsep=0pt, leftmargin=1cm]
    \item {\color{color1}$\tilde{p}_i$} represents the worst-case mandatory hospital stay for patient $i$, encompassing both the baseline clinical recovery requirements and the maximum anticipated operational delays (e.g., staffing shortages or administrative bottlenecks), 
    \item {\color{color1}$E_i$} denotes a binary variable indicating whether patient $i$ stayed longer than $\tilde{p}_i$, where $E_i = 1$ if the stay exceeded $\tilde{p}_i$ and $E_i = 0$ otherwise,
    \item {\color{color1}$P_i \in[0,\tilde{p}_i]$} denotes the actual protocol-driven stay realized by patient $i$. This duration encompasses the baseline clinical requirement plus any experienced operational delays, and is bounded by the maximum threshold,
    \item {\color{color1}$y_{_{E_i}}\in[0,u-\tilde{p}_i]$} denotes the duration of the (extended) stay patient $i$ spent in hospital after surgery beyond the worst-case protocol time $\tilde{p}_i$.
    \end{itemize}
\item {\color{color2}$y_{_{S_i}} = y_{_{C_i}}\mathbbm{1}(C_i = 1)$} denotes the total subsequent days patient $i$ spent in post-discharge care (hospital, respite, or nursing facilities, depending on the selected DAH definition), provided these facilities were not their usual residence at baseline, where
\begin{itemize}[itemsep=-5pt,topsep=0pt, leftmargin=1cm]
    \item {\color{color2}$C_i$} denotes a binary variable indicating whether patient $i$ required post-discharge care, where $C_i = 1$ if care was required and $C_i = 0$ otherwise,
    \item ${\color{color2}y_{_{C_i}}}\in[0,u-{\color{black}y_{_{I_i}}}]$ denotes the duration of post-discharge care for patient $i$ if extra care is required.
    \end{itemize}
\item {\color{color3}${\color{color3}\mathbbm{1}(D_i = 0)}$} denotes a binary variable indicating the death status of patient $i$, where $D_{i}$~=~1 if the patient $i$ died over the period of interest and $D_{i}$ = 0 if the patient was alive on day $u$.
\end{itemize}

\par\noindent In the following, we describe candidate statistical models for the different components defined in Equation (1). We exclude the deterministic elements of the equation, specifically:
\begin{itemize}[nosep, leftmargin=1cm]
    \item $u$, the duration of the follow-up window, which is fixed and pre-defined,
    \item $\tilde{p}_i$, the maximum protocol stay given patient characteristics, which is considered as known and informed by clinical expertise.\\ 
\end{itemize}

\par\noindent \textit{Death Status}, ${\color{color3}\mathbbm{1}(D_i = 0)}$\\
The binary mortality status of patient $i$ is modeled via logistic regression, linking the probability of death ($\pi_{d_i}$) to predictors using a logit link, i.e.,
\begin{equation}
P(D_i=1)=\pi_{d_i}=\frac{\exp\big(x_{{\scriptscriptstyle D}_i}^{T}\beta{{\scriptscriptstyle D}}\big)}{1+\exp\big(x_{{\scriptscriptstyle D}_i}^{T}\beta{_{\scriptscriptstyle D}}\big)}
\label{eq:model}
\end{equation}
where $x_{{\scriptscriptstyle D}_i}$ and $\beta{{\scriptscriptstyle D}}$ respectively denote the covariate vector for participant $i$ and the corresponding vector of regression coefficients related to survival.\\

\par\noindent \textit{Length of Postoperative Hospital Stay}, {\color{color1}$y_{_{I_i}}$}\\
The different elements composing this random variable can be modeled as follows:
\begin{itemize}[itemsep=-5pt,topsep=0pt, leftmargin=1cm]
    \item \textit{Extended Initial Hospital Stay Status}, ${\color{color1}\mathbbm{1}(E_i = 1)}$\\ 
    Similarly as in (\ref{eq:model}), the binary indicator for whether patient $i$ requires an extended hospital stay can be modeled via logistic regression, which links the probability of an extended stay for patient $i$, $P(E_i=1) = \pi_{_{E_i}}$, to patient-related predictors, $x_{{\scriptscriptstyle E_1}_i}$, and corresponding regression coefficient vector, $\beta_{{\scriptscriptstyle E_1}}$, via the logit link function.
    \item \textit{Length of Postoperative Protocol Stay}, ${\color{color1}P_i}$ \\
    Due to operational frictions, the protocol-driven stay for patient $i$ is treated as a discrete random variable that can be modeled via a multinomial or ordinal regression model, which assigns a probability to each possible duration, $P(P_i = x)$ for $x \in \{0, \dots, \tilde{p}_i\}$. These probabilities may optionally be linked to patient-specific predictors $x_{{\scriptscriptstyle P}_i}$, such as country and hospital, and its corresponding regression coefficient vector, $\beta_{{\scriptscriptstyle P}}$, through a suitable link function (e.g., a cumulative logit link for ordinal data).
    \item \textit{Length of Postoperative Extended Stay}, ${\color{color1}Y_{_{E_i}}}$\\
    The duration of the extended postoperative stay, $\color{color1}y_{_{E_i}}$, can be modeled using a zero-truncated, right-censored count distribution, such as the zero-truncated and right-censored variants of the Poisson, negative binomial, or Poisson inverse Gaussian. The latter two distributions, respectively representing Gamma and Inverse Gaussian mixtures of Poisson variates, are particularly well-suited for handling over-dispersed count data. Since the domain of $\color{color1}y_{_{E_i}}$ is bounded by $[0, u - \tilde{p}_i]$, a right-censored framework is essential, especially when the observation window $u$ is short. In this context, right-censored observations result in a DAH values of zero and constitute a second source of zeros in the model, distinct from those induced by death.\\
    We use the Generalized Additive Models for Location, Scale, and Shape (GAMLSS) framework \citep{stasinopoulos2017} to connect the parameters of the selected distribution (e.g., mean and dispersion) to patient-specific predictors. GAMLSS extends the Generalized Linear Model (GLM) by allowing all distributional parameters to vary with covariates. Using a log link function, the expected duration of the extended stay for patient $i$ could then be expressed as $\mu_{_{E_i}} = \exp(x_{{\scriptscriptstyle E_{\mu}}_i}^T\beta_{{\scriptscriptstyle E_{\mu}}})$, where $x_{{\scriptscriptstyle E_{\mu}}_i}$ denotes the vector of patient-specific predictors and $\beta_{{\scriptscriptstyle E_{\mu}}}$ represents the corresponding vector of regression coefficients.
\end{itemize}
Note that this joint modeling of $\color{color1}E_i$ and $\color{color1}y_{_{E_i}}$ operates as a zero-adjusted (hurdle) model, which strictly separates the processes: the logistic component generates all zeros, while the zero-truncated, right-censored count distribution generates strictly positive values. An alternative joint approach  would be to consider a zero-inflated model, where the distribution of $\color{color1}y_{_{E_i}}$ is not zero-truncated. In this case, the logistic component models the `zero-inflation' probability, i.e., the proportion of excess zeros occurring in addition to the zeros that naturally arise from the right-censored count distribution \citep[see][for details]{min2005}.\\

\par\noindent \textit{Length of Post-Discharge Care}, {\color{color2}$Y_{_{S_i}}$}\\
The duration of post-discharge care for patient $i$, denoted by $y_{_{S_i}}$, is bounded between $0$ (if no care is required after the first discharge) and $u - y_{_{I_i}}$ (if the patient requires extra care for the entire remaining observation window). While various modeling strategies exist, including the zero-adjusted or zero-inflated right-censored count distributions discussed for $\color{color1}y_{_{E_i}}$, we focus here on a zero-adjusted beta-binomial model. This approach models the proportion of the post-discharge period spent in extra care and naturally accounts for the patient-specific time window through the binomial denominator. The two elements composing this random variable can be modeled as follows:
\begin{itemize}[itemsep=-5pt,topsep=0pt, leftmargin=1cm]
    \item \textit{Post-Discharge Care Status}, ${\color{color2}\mathbbm{1}(C_i = 1)}$\\ 
    Similarly as in (\ref{eq:model}), the binary indicator for whether patient $i$ requires post hospital discharge care can be modeled via a logistic regression, which links the probability of requiring post discharge care for patient $i$, $P(C_i=1) = \pi_{_{C_i}}$, to patient-related predictors, $x_{{\scriptscriptstyle C_1}_i}$, and corresponding regression coefficient vector, $\beta{{\scriptscriptstyle C_1}}$, via the logit link function.
    \item \textit{Fraction of Time Spent in Post-Discharge Care}, ${\color{color2}Y_{_{C_i}}}$\\
    A zero-truncated variant of the binomial or beta-binomial distribution allows for modeling the number of days spent in post-discharge care as a sequence of `successes' out of the $n_i = u - y_{_{I_i}}$ patient-specific available days. By assuming that the probability of success is beta distributed, the beta-binomial model generalizes the binomial distribution to effectively capture overdispersion and participant-level heterogeneity. \\
    The expected proportion of post-discharge care for patient $i$, $\mu_{_{C_i}}$, is linked to predictors $x_{{\scriptscriptstyle C_2}_i}$ and coefficients $\beta{{\scriptscriptstyle C_2}}$ via a logit link function. Within the GAMLSS framework, the dispersion parameter of the beta-binomial distribution can also be modeled as a function of covariates using an appropriate link function.
\end{itemize}

Note that, similarly to the modeling of $\color{color1}y_{_{E_i}}$, a zero-inflated variant of the beta-binomial distribution may be employed. In this framework, the observed zeros result from a mixture of two distinct processes: `structural' zeros, representing patients who are essentially excluded from requiring extra care, and `sampling' zeros, which arise within the beta-binomial process for patients who remained at their baseline level of care despite being at risk for post-discharge requirements.

If observations are dependent -- as a consequence of a clustering by region or hospital, for example -- the models described above can be extended to include (possibly correlated) random effects in the different parts of the models to take that dependence into account.

Of note, other commonly used composite endpoints like the number of ``Days Alive and Out of Hospital'' and the number ``Ventilator Free Days'' can be expressed as special cases of our general model.

\section{Modeling the NOTACS Data}
\label{s:notacs-application}
We fit the divide and conquer model to the data of 200 patients from the NOTACS trial interim analysis displayed in Figures \ref{fig:fig1} and \ref{fig:fig2} in Section \ref{ss:notacs-application} and assess its fit, by part and overall, in Section \ref{sec:modelcheck}. Compared to competing models, the inherent flexibility of our approach leads to a demonstrably superior overall fit.

\subsection{Model fit}
\label{ss:notacs-application}

Without loss of generality, we opted for the following specifications:
\begin{itemize}
    \item based on clinical input, $\tilde{p}$ was set to 4 days in the 3 countries. As no patient was discharged before day 4, the model for $P_i$ was redundant, and our model for $y_{_{I_i}}$ simplified to $${\color{color1}y_{_{I_i}} = \tilde{p}^{\,{\mathbbm{1}(E_i = 0)}} [\tilde{p} + y_{_{E_i}}]^{{\mathbbm{1}(E_i = 1)}}}$$
    \item By equating subsequent care ($\color{color2}y_{S_i}$) solely with hospital readmissions, we effectively analyzed 'Days Out of Hospital' (DOOH). This substitution minimally impacted overall values, as only 1\% of patients ($n=2$) required respite care (Upper panel of Figure \ref{fig:fig1}, patients 30 and 166).
    \item Because only 2 patients died, the model part related to ${\color{color3}\mathbbm{1}(D_i = 0)}$, the death status, did not consider covariates.
    \item We considered the following predictors (as well as their interactions) to model the parameters related to $\color{color1}y_{_{E_i}}$ and $\color{color2}y_{_{C_i}}$: sex, (intention-to-treat) treatment allocation, body mass index (classified relative to the median), age group ($\leq$50, $>$50) and country. As the readmission length may be associated with the length of the initial hospital stay \citep{bueno2010,carey2014}, we additionally considered $\color{color1}y_{_{E_i}}$ and $\color{color1}\mathbbm{1}(y_{_{E_i}} = 0)$, the extended initial hospital stay and corresponding status, as a predictors of parameters of $\color{color2}y_{_{S_i}}$, the subsequent care. This allows the linking of both model parts.
\end{itemize}

We used the GAMLSS framework, as implemented in the   \texttt{gamlss} (version 5.4.22) and \texttt{gamlss.cens} (version 5.0.7) R packages as follows. For each model part (i.e., $\color{color1}y_{_{E_i}}$, $\color{color2}y_{_{C_i}}$), several candidate conditional distributions were considered and a variable selection was performed following the stepwise selection procedure described by \citeauthor{stasinopoulos2017} (\citeyear[Chapter 11.5]{stasinopoulos2017}). Starting from a null model, forward variable selection based on the Generalized Akaike Information Criterion (GAIC) was performed sequentially for the location ($\mu$), scale ($\sigma$), and clump-at-zero ($\nu$) parameters. This process was then followed by backward elimination to identify the `optimal' predictor set for each parameter of each conditional distribution. The final model of each model part was defined as the candidate distribution and parameter-related set of predictors leading to the smallest GAIC. This process led us to select the following conditional distributions and model predictors:
\begin{itemize}
    \item For ${\color{color3}\mathbbm{1}(D_i = 0)}$, a logistic model without covariate (as explained above),
    \item For ${\color{color1}y_{_{E_i}}}$, a zero-inflated Poisson-inverse Gaussian (PIG) model with right censoring (ZICPIG),  
    \begin{equation}
    {\color{color1}y_{_{E_i}} \sim \text{ZICPIG}(\pi_{_{E_i}}, \mu_{_{E_i}}, \phi_{_{E_i}}, u - \tilde{p})}    
    \end{equation}
    with the following predictors per parameter:
    \begin{itemize}
        \item ${\color{color1}\pi_{_{E_i}}}$, the probability of excess zeroes, was linked to no covariate,
        \item ${\color{color1}\mu_{_{E_i}}}$, the average of the extended stay, assumed PIG distributed with right censoring, for patients whose initial stay exceeds the minimum duration, was linked to `country`, `sex', `age', `treatment', `BMI' (with an interaction between `treatment' and `BMI') via a log link function, 
        \item ${\color{color1}\phi_{_{E_i}}}$, the shape parameter of the extended stay, was linked to to the same variables as $\mu_{_{E_i}}$ expect `age' via a log link function, 
    \end{itemize}
    \item For ${\color{color2}y_{_{C_i}}}$, a zero-adjusted beta-binomial model (ZABB) 
    \begin{equation}
    {\color{color2}y_{C_i} \sim \text{ZABB}(\pi_{_{C_i}}, \mu_{_{C_i}}, \phi_{_{C_i}}, u - \tilde{p} - y_{_{E_i}})}
    \end{equation}
    the following predictors per parameters:
    \begin{itemize}
        \item ${\color{color2}\pi_{_{C_i}}}$, the probability of hospital readmission, was linked to $\color{color1}\mathbbm{1}(y_{_{E_i}} = 0)$ and `country' via a logit link,        
        \item ${\color{color2}\mu_{_{C_i}}}$, the fraction of hospital stay due to post-discharge readmission, assumed to be beta-binomial (BB) distributed with patient binomial denominators set to $u - \tilde{p} - y_{_{E_i}}$, was linked to $\color{color1}y_{_{E_i}}$ via a logit link,
        \item ${\color{color2}\phi_{_{C_i}}}$, the BB shape parameter, was linked to `treatment' and $\color{color1}\mathbbm{1}(y_{_{E_i}} = 0)$ via a log link.        
    \end{itemize}    
\end{itemize}

Our full model parameter vector is given by
$\bm{\theta} = \left[  {\color{color3}\pi_{d_i}}, {\color{color1}\pi_{e_i}}, {\color{color1}\mu_{e_i}}, {\color{color1}\phi_{e_i}}, {\color{color2}\pi_{_{C_i}}}, {\color{color2}\mu_{_{C_i}}}, {\color{color2}\phi_{_{C_i}}} \right]^T$. Table~\ref{fig:fitestimates} reports the $\beta$ regression parameter estimate, corresponding $p$-value and significance level for each selected predictor of each model part. Note, for example, that country has an effect on both the length of the initial hospital stay and the probability of readmission, and that the readmission length is positively associated with the length of the initial hospital stay conditionally on other predictors. 

\begin{table}[htbp]
\caption{$\beta$ regression parameter estimate, corresponding p-value and significance level for each selected predictor of each model part (n=200).}
\centering
\fontsize{9}{14}\selectfont
\setlength{\tabcolsep}{4pt}
\renewcommand{\arraystretch}{0.95}

\begin{tabular}{@{}p{3cm}>{\centering\arraybackslash}p{1.6cm}>{\centering\arraybackslash}p{1.6cm}>{\centering\arraybackslash}p{1.6cm}>{\centering\arraybackslash}p{1.6cm}>{\centering\arraybackslash}p{1.6cm}>{\centering\arraybackslash}p{1.6cm}>{\centering\arraybackslash}p{1.6cm}@{}}
\toprule
\textbf{Predictor} 
& $\boldsymbol{{\color{color3}\pi_{_{D}}}}$ 
& $\boldsymbol{{\color{color1}\pi_{_{E_i}}}}$ 
& $\boldsymbol{{\color{color1}\mu_{_{E_i}}}}$ 
& $\boldsymbol{{\color{color1}\phi_{_{E_i}}}}$ 
& $\boldsymbol{{\color{color2}\pi_{_{C_i}}}}$ 
& $\boldsymbol{{\color{color2}\mu_{_{C_i}}}}$ 
& $\boldsymbol{{\color{color2}\phi_{_{C_i}}}}$ \\[1pt]

\textbf{Link function}
& logit
& logit
& log
& log
& logit
& logit
& log \\[1pt]

\midrule

\textit{Intercept}
& \estp{-4.595}{($<0.001$***)}
& \estp{-36.040}{(0.996)}
& \estp{0.935}{(0.026*)}
& \estp{1.475}{($<0.001$***)}
& \estp{2.762}{($<0.001$***)}
& \estp{-2.917}{($<0.001$***)}
& \estp{-36.075}{($<0.001$***)} \\[1pt]

age groups
&
&
& \estp{0.939}{(0.012*)}
&
&
&
& \\[1pt]

treatment
&
&
& \estp{-1.236}{($<0.001$***)}
& \estp{-1.417}{(0.013*)}
&
&
& \estp{-1.831}{(0.029*)} \\[1pt]

BMI
&
&
& \estp{-0.629}{(0.029*)}
& \estp{-1.088}{(0.039*)}
&
&
& \\[1pt]

sex
&
&
& \estp{1.166}{($<0.001$***)}
& \estp{1.134}{(0.005**)}
&
&
& \\[1pt]

\shortstack[l]{country: Australia}
&
&
& \estp{1.136}{(0.001***)}
& \estp{0.225}{(0.702)}
& \estp{-0.412}{(0.410)}
&
& \\[1pt]

\shortstack[l]{country: New Zealand}
&
&
& \estp{0.192}{(0.422)}
& \estp{-1.530}{(0.003**)}
& \estp{-1.120}{(0.009**)}
&
& \\[1pt]

BMI $\times$ treatment
&
&
& \estp{1.733}{($<0.001$***)}
& \estp{2.311}{(0.004**)}
&
&
& \\[1pt]

sex $\times$ treatment
& \estp{~}{~}
& \estp{~}{~}
& \estp{~}{~}
& \estp{~}{~}
& \estp{~}{~}
& \estp{~}{~}
& \estp{~}{~} \\[1pt]

age $\times$ treatment
& \estp{~}{~}
& \estp{~}{~}
& \estp{~}{~}
& \estp{~}{~}
& \estp{~}{~}
& \estp{~}{~}
& \estp{~}{~} \\[1pt]

$y_{E_i}$
&
&
&
&
&
& \estp{0.070}{($<0.001$***)}
& \\[1pt]

$\mathbbm{1}(y_{E_i}=0)$
&
&
&
&
& \estp{-1.427}{(0.062)}
&
& \estp{34.785}{($<0.001$***)} \\

\bottomrule
\end{tabular}

\vspace{2mm}
\parbox{0.95\linewidth}{\fontsize{9.5}{11}\selectfont *** $p<0.001$, ** $p<0.01$, * $p<0.05$.}
\label{fig:fitestimates}
\end{table}

\subsection{Model Evaluation}
\label{sec:modelcheck}
We assess the fit of our ‘Divide \& Conquer’ model at both the individual component and global composite levels, followed by a direct comparison with traditional parametric models to verify its performance.

\subsubsection{Model check per model part}
\label{sec:modelcheck_part}
For each model part, the adequacy of the selected model was assessed using the `Worm' plots displayed in Figure \ref{fig:rqres} which compare the normalized randomized quantile residuals \citep{dunn1996} and their (approximate) expected values represented by the horizontal dotted line. These model checks suggest a good fit of the selected model for each model part as the vast majority of the randomized residuals lie within the approximate point-wise 95\% confidence intervals (elliptic curves), indicating no evidence against the selected models \citep[see][Chapter 12.4, for details]{stasinopoulos2017}.

\subsubsection{Overall model}
\label{sec:modelcheck_overall}
Alternatively, a model check can be performed at the composite-endpoint level to assess the adequacy of the fitted model in reproducing the marginal distribution of DAH, using a resampling-based predictive check as follows. For a given model:
\begin{itemize}
   \item First, we generate $B=5{,}000$ resampled and synthetic datasets as follows. For each replicate, we randomly draw $n$ observed data (pairs of DAH values and corresponding predictor vectors) with replacement to obtain a nonparametric bootstrap sample, and then simulate $n$ new DAH values from the fitted model using the same resampled predictor design matrix. This process results in $B$ pairs of resampled and synthetic datasets, each of size $n$. We then compute empirical quantiles from the resampled DAH values and model-based quantiles from the simulated values at each point of a grid of $250$ equally spaced target probabilities $p_k \in (0, 1)$, $k = 1, \ldots, 250$.  
    \item Second, a resampling-based quantile--quantile predictive model check plot is constructed as follows. For each $p_k$, the $x$-coordinate is the bootstrap mean of the empirical quantile across replicates, and the $y$-coordinate is the bootstrap mean of the model-based quantile. Around each $y$-coordinate, an envelope is drawn between the pointwise $2.5\%$ and $97.5\%$ bootstrap quantiles of the model-based quantile distribution, reflecting both the sampling variability of the predictor distribution and the stochastic variability induced by the fitted model. 
\end{itemize}

This diagnostic assesses the model's ability to reproduce the marginal distribution of DAH. A well-specified model is indicated by alignment of the paired mean points along the identity line, with the envelope covering the identity line across the range of $p_k$.

The upper-left panel of Figure \ref{fig:modelcheck_resampling} shows the output of this model check approach applied to the `Divide \& Conquer' fit of the NOTACS interim data presented in Table \ref{fig:fitestimates}. This plot shows a remarkable alignment of the observed and simulated quantiles over the entire domain of interest. The mean of the quantiles simulated under the model closely matches the identity line, which remains centered within the band constructed from the 0.025 and 0.975 quantiles of the simulated DAH data. The bandwidth reflects the density of the data: extremely narrow for DAH values greater than 75, where observations are concentrated, and wide for values below 50, where observations are sparser. This model check thus suggests an excellent fit of the `Divide \& Conquer' model to the data.

\subsubsection{Comparative evaluation of the fit} 
\label{sec:ComparativeFitEval}

We compared our approach against several existing parametric models (Table~\ref{tab:existingmodel}), fitting each via the GAMLSS framework as described above to ensure consistency. Each comparator was adapted to match our DAH definition; for instance, the negative binomial model \citep{vanhoutven2019} was extended with a logistic component to handle mortality-related zeros. For models with underspecified parameters, we applied optimized configurations (see Table~\ref{tab:existingmodel}) to facilitate a fair and robust performance benchmark. These adjustments ensure that the comparative analysis reflects each model's maximum potential fit within the context of the NOTACS data.

\begin{figure}[!htbp]
    \includegraphics[width=0.7\textwidth, keepaspectratio]
    {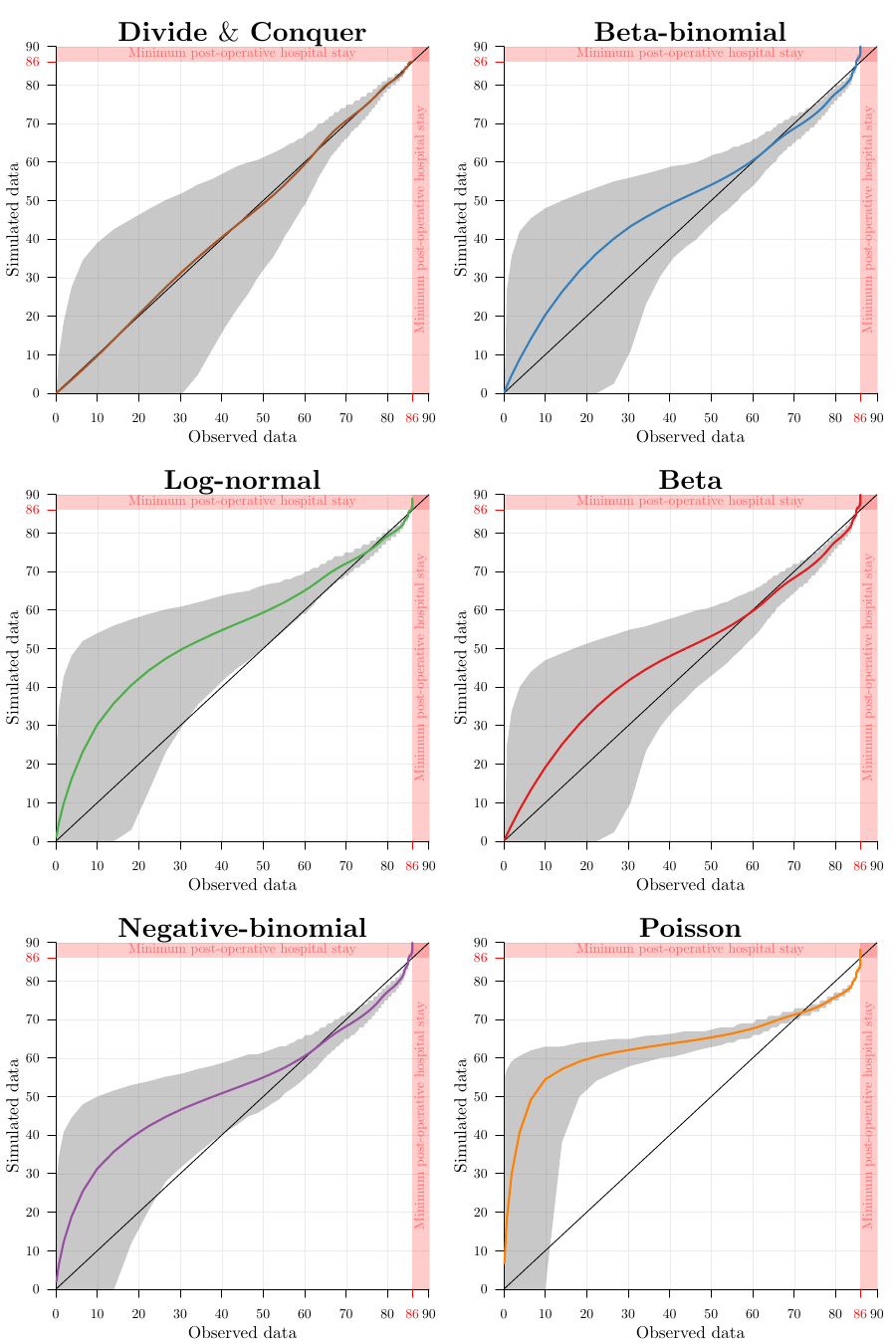}
    \centering
    \caption{Resample-based quantile--quantile predictive check for the
    fitted DAH model: Axes show DAH quantiles in days, evaluated at $250$
  equally spaced probability points in $(0, 1)$. The x-axis gives the
  bootstrap mean of empirical quantiles across $B$ resamples of the
  observed data, and the y-axis gives the bootstrap mean of quantiles
  from data simulated under the fitted model using the resampled
  predictor matrix. The colored line connects these paired points across
  the probability grid; shaded bands give pointwise $95\%$ bootstrap
  intervals ($2.5\%$--$97.5\%$ across replicates) of the model-simulated
  quantiles. The black diagonal line is the identity; alignment of the colored
  line with the identity, within the envelope, indicates that the model
  adequately reproduces the marginal distribution of DAH. The minimum hospital stay, $\tilde{p}$, equals 4 days for the `Divide \& Conquer' model and 0 for other models.}

    \label{fig:modelcheck_resampling}
\end{figure}

Figure \ref{fig:modelcheck_resampling} shows the results of the overall model check approach described in Section \ref{sec:modelcheck_overall}, applied to the NOTACS interim data for each methodology. We observe that the poorest fits correspond to models that assume (flipped) Poisson, negative binomial, or log-normal distributions for strictly positive DAH values. While models assuming a beta or beta-binomial distribution obtained better diagnostic results, they failed to match the quality of the fit achieved by the `Divide \& Conquer' approach.
A potential reason for these poor fits is the omission of the minimum post-intervention stay (i.e., $\tilde{p} = 4$ in our dataset), which allows for DAH values greater than $u-\tilde{p} = 86$ when considering competing modeling. To address this, we refitted each competing method by subtracting $\tilde{p}$ from the positive DAH values. While the resulting model checks, available in Figure \ref{fig:modelcheckB-bootstrap},  show an improved fit across all models -- highlighting the importance of this parameter -- none reached the quality of fit obtained by our proposed approach. This conclusion is further supported by Figure \ref{fig:discrepancy}, which compares the performance of each model by reporting the integrated discrepancy between its Q--Q curve, displayed
in Figure \ref{fig:modelcheck_resampling} and \ref{fig:modelcheckB-bootstrap}, and the identity line (i.e., the area enclosed between the two), showing that the `Divide \& Conquer' model achieves the best performance.

In addition to considering a minimum post-intervention stay, we believe the superior fit of the `Divide \& Conquer' model comes from its ability to separately model the initial postoperative stay, $y_{I{_i}}$, and subsequent care, $y_{S{_i}}$, whereas competing methods implicitly aggregate these components by considering their sum. Being unimodal, these approaches are unable to account for the mixture of DAH values solely defined by the first component and those defined by both, thereby leading to a poorer fit.

\clearpage

\section{Beyond Analysis: Enhancing Trial Design Through High-Fidelity DAH Simulation}
\label{sec:modelapplication}

The benefit of a superior modeling approach extends from retrospective trial analysis to prospective trial design. Having established a model strategy capable of generating realistic DAH data, we can now evaluate the operating characteristics of various trial designs under finite-sample conditions. This is a crucial step that goes beyond the common practice of simply selecting a test, such as the Mann-Whitney-Wilcoxon (MWW) \citep{myles2018, wong2022, notacs2026}, quantile regression \citep{myles2017} or t-test \citep{chung2023}, by allowing investigators to quantify the reliability of their proposed methodology before implementation.

In this Section, we thus investigate the operating characteristics of the MWW test, as it was the primary analysis method used to compare the two arms of the NOTACS trial \citep{notacs2026} which motivated this work. Via a Monte Carlo simulation, we thus evaluate the probability of rejecting the null hypothesis under two following scenarios:

First, we treated a subset of the NOTACS interim data ($n=200$) as a pilot dataset, enabling us to estimate the
`Divide \& Conquer' model parameters for the control arm: the probability of death ${\color{color3}\pi_{d_0}}$, the probability of exceeding the minimum postoperative hospital stay ${\color{color1}\pi_{e_0}}$, the extended initial stay average and shape parameters ${\color{color1}\mu_{e_0}}$ and ${\color{color1}\phi_{e_0}}$, the probability of post-discharge subsequent care ${\color{color2}\pi_{_{C_0}}}$, and the average and nuisance parameters of the fraction of time spent in subsequent care ${\color{color2}\mu_{_{C_0}}}$ and ${\color{color2}\phi_{_{C_0}}}$. 
By fitting the model to this representative sample, we obtained high-fidelity parameter estimates that capture the complex, zero-inflated, and bounded nature of the DAH distribution. No additional predictors were included in the model, as the MWW test exclusively evaluates the treatment effect.

Second, using these estimated parameters, we defined realistic null and alternative scenarios. The null scenario assumes no treatment effect between arms. For the alternative scenario, while the treatment effect could theoretically influence any of the model components, clinical input suggested that HFNT would primarily impact the average extended initial stay, ${\color{color1}\mu_{e_0}}$. We therefore used simulation to identify the range of treatment-related log-fold change values that produce a between-group median difference of two days, consistent with the NOTACS trial protocol, and selected the midpoint of this range, as illustrated in Figure \ref{fig:beta1-stepfunction}. 
For comparative purposes, we repeated this calibration for each competing model listed in Table~\ref{tab:existingmodel}, applying the same adaptations detailed in Section~\ref{sec:ComparativeFitEval}.

Third, we performed a Monte Carlo simulation over a grid of sample sizes ranging from $n = 100$ (50 per arm) to $n = 2{,}000$ (1{,}000 per arm), assuming a two-sided test and a 5\% type~I error threshold. For each $n$, we generated 10{,}000 trial realizations under each scenario and computed the empirical rejection rate, yielding the statistical power under the alternative scenario and the Type~I error rate under the null. 
This allowed us to identify the minimum sample size required to achieve 90\% power while verifying that the Type~I error rate of the MWW test remained controlled at the nominal level across all data-generating processes considered.

\begin{figure}[!htbp]
\centering
\includegraphics[width=0.8\textwidth]{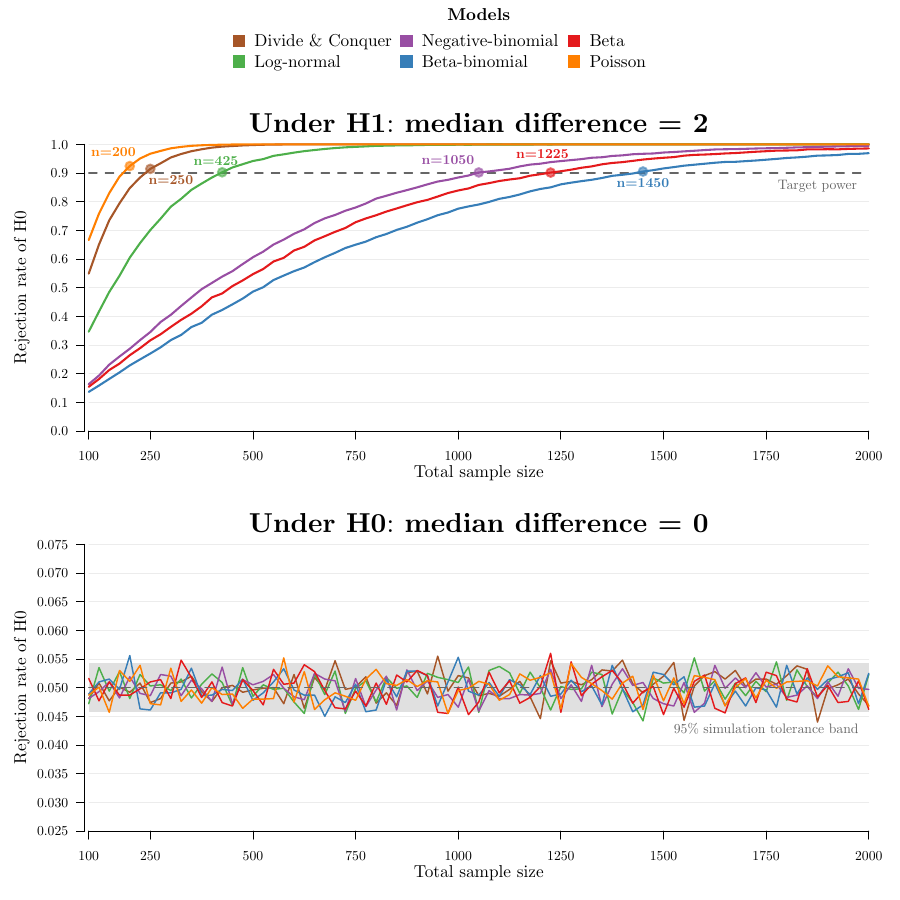}
\caption{Upper plot: Probability of rejecting the null hypothesis (y-axis) under the alternative hypothesis (when assuming a between-group median difference of two) as a function of the total sample size (x-axis) for each method (colored lines). Lower plot: same when the null hypothesis is true (i.e., when assuming a between-group median difference of 0). The gray rectangle shows the Monte Carlo pointwise 95\% simulation tolerance band.}
\label{fig:power-null}
\end{figure}

Figure \ref{fig:power-null} presents the simulation results. The upper panel shows the power of the MWW test under the settings described above as a function of the sample size for each method. While a sample size of 250 patients (125 per arm) appears sufficient to detect the targeted treatment effect under the `Divide \& Conquer' model, the sample size required by other models to achieve the same power varies substantially, ranging from 200 to 1{,}450 patients. The wide variation highlights how a poorly fitting model can easily lead to over- or underestimation of sample size, resulting in inefficient trial designs or unreliable results. Note that, under the alternative, the treatment effect modifies the average extended initial stay, ${\color{color1}\mu_{_{E_i}}}$, which induces differences not only in central tendency but also in shape and dispersion between the two arms. The behavior of the MWW test is known to be sensitive to such differences \citep{fagerland2009}, making its power under these data-generating processes difficult to anticipate analytically and motivating the simulation-based approach adopted here.

The lower panel of Figure~\ref{fig:power-null} shows the empirical Type~I error rate under the same settings. We can note that the MWW test maintains the expected nominal size of 5\% across all modeling frameworks considered, confirming that the test is correctly calibrated under each of the data-generating processes used in the power calculation.

\clearpage

\section{Discussion}
\label{s:discuss}

The `Divide \& Conquer’ framework offers a shift from treating complex outcomes like DAH as single, collapsed metrics to handling them as clinically distinct components. This approach provides three primary advantages for trial design.

First, statistical fidelity is a prerequisite for ethical trial design. Traditional models that ignore the zero-inflated, bounded nature of DAH often misspecify the data-generating process, leading to inaccurate sample size calculations. This introduces the dual risk of underpowered trials that fail to detect meaningful effects and overpowered trials that unnecessarily expose patients to experimental risks. By accurately modeling the underlying distribution, our framework ensures trials are both statistically robust and ethically sound.

Second, the framework enhances clinical interpretability. While rank-based tests like the MWW collapse all information into a single metric, our approach preserves the clinical drivers of treatment effect. Investigators can explicitly incorporate prior expectations of how a treatment is expected to act, e.g., a reduction in post-discharge readmissions versus a shift in mortality, and let these insights guide both the design and analysis. This granular understanding ensures that the trial's statistical power is optimized for the specific clinical pathway the intervention targets, directly aligning analysis with real-world clinical mechanisms.

Finally, while this framework is more computationally intensive than standard tests, the trade-off is favorable. The modest resource cost required for model fitting and Monte Carlo simulation at the design stage is negligible compared to the financial and human costs of a failed 1{,}000-patient trial resulting from a misspecified design.

 The framework currently requires component-level data from pilot studies or electronic health records to inform simulations. Future iterations will adapt the model for shorter windows (e.g., DAH30) and more complex missing data mechanisms \citep{Tackney2026}. Furthermore, the inherent limitation of treating mortality and prolonged hospitalization as mathematically similar on the DAH scale warrants further investigation into the DAH estimand. Given its flexibility, we see significant potential for this approach in adaptive designs for sample size re-estimation and in other bounded composite outcomes, such as ventilator-free days.

 By decomposing DAH into mortality, initial stay, and post-discharge care, our framework aligns with current FDA recommendations for complex endpoints \citep{fda2019}. It provides a high-fidelity blueprint for trialists to move beyond simplistic parametric assumptions, ensuring that clinical trial designs are both statistically rigorous, ethically sound and clinically grounded.

\backmatter


\section*{Acknowledgments}

LY is supported by the MRC Trial Methodology Research Partnership Doctoral Training Program (TMRP DTP, grant number MR/W006049/1). DLC and SSV are partly funded by the UK MRC grant number MC\_UU\_00002/1.\vspace*{-8pt}


%


\section*{Supplementary Materials}

Additional figures referenced in Sections~
\ref{s:notacs-application} and~\ref{sec:modelapplication} are provided below.
\vspace*{-8pt}

\clearpage

\setcounter{figure}{0}
\setcounter{table}{0}

\renewcommand{\thefigure}{S.\arabic{figure}}
\renewcommand{\thetable}{S.\arabic{table}}

\definecolor{lightviolet}{rgb}{0.97, 0.95, 1.0}

\definecolor{color1}{HTML}{FF7F50}
\definecolor{color2}{HTML}{6610F2}
\definecolor{color3}{HTML}{4CBB17}




\begin{figure}[ht]
\centering

{\normalsize\bfseries {\color{color3}$\mathbbm{1}(D_i = 0)$}\par}
\vspace{-24pt}
\includegraphics[width=0.5\textwidth]{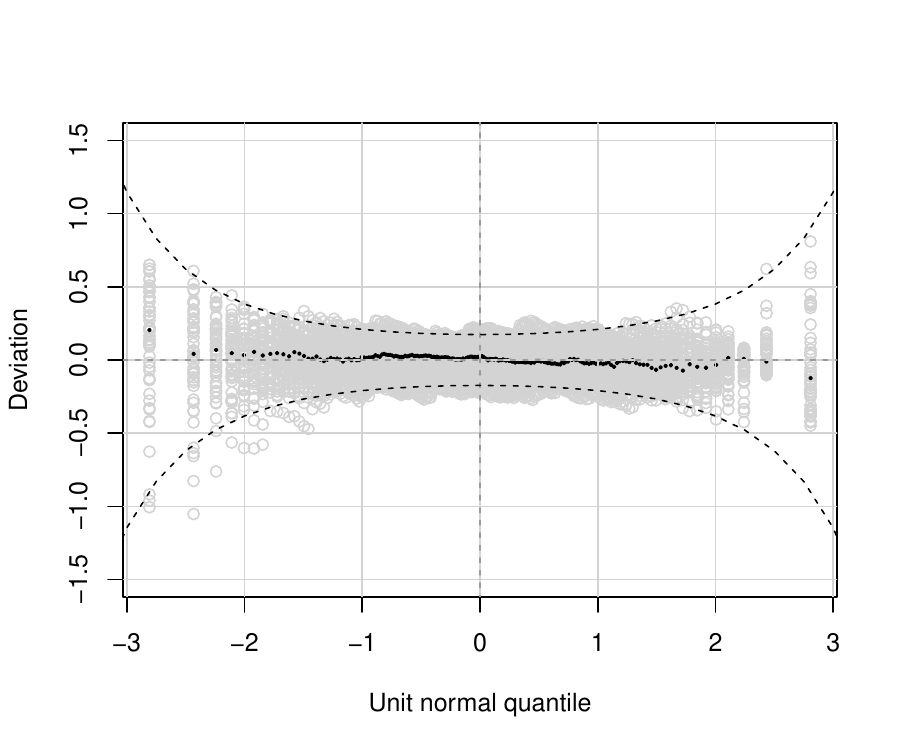}

\vspace{1pt}
{\normalsize\bfseries {\color{color1}$y_{_{I_i}}$}\par}
\vspace{-24pt}
\includegraphics[width=0.5\textwidth]{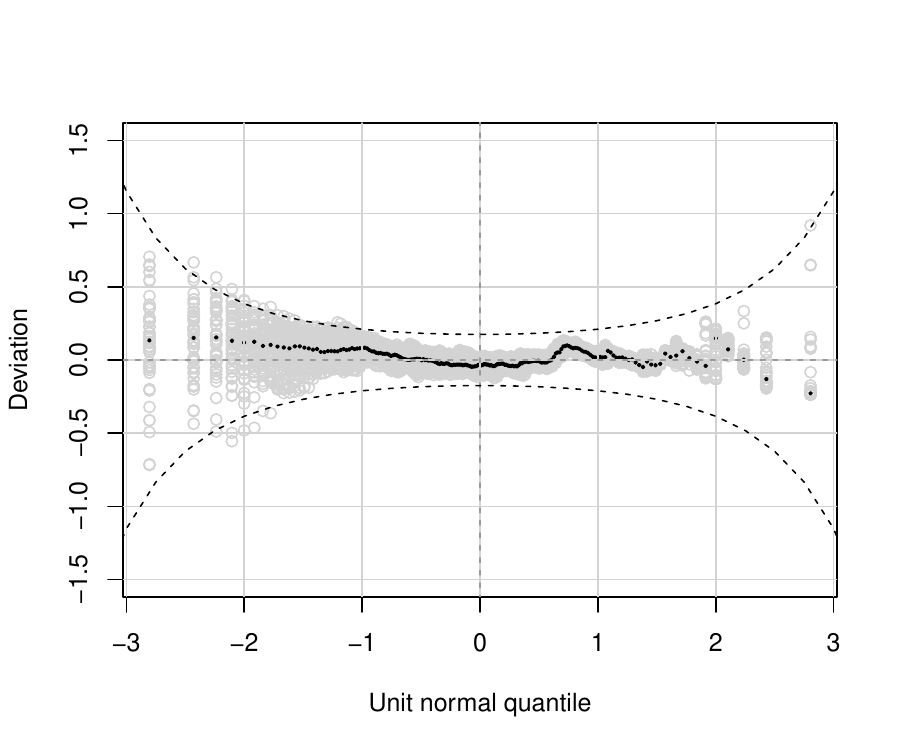}

\vspace{1pt}

{\normalsize\bfseries {\color{color2}$y_{_{S_i}}$}\par}
\vspace{-24pt}
\includegraphics[width=0.5\textwidth]{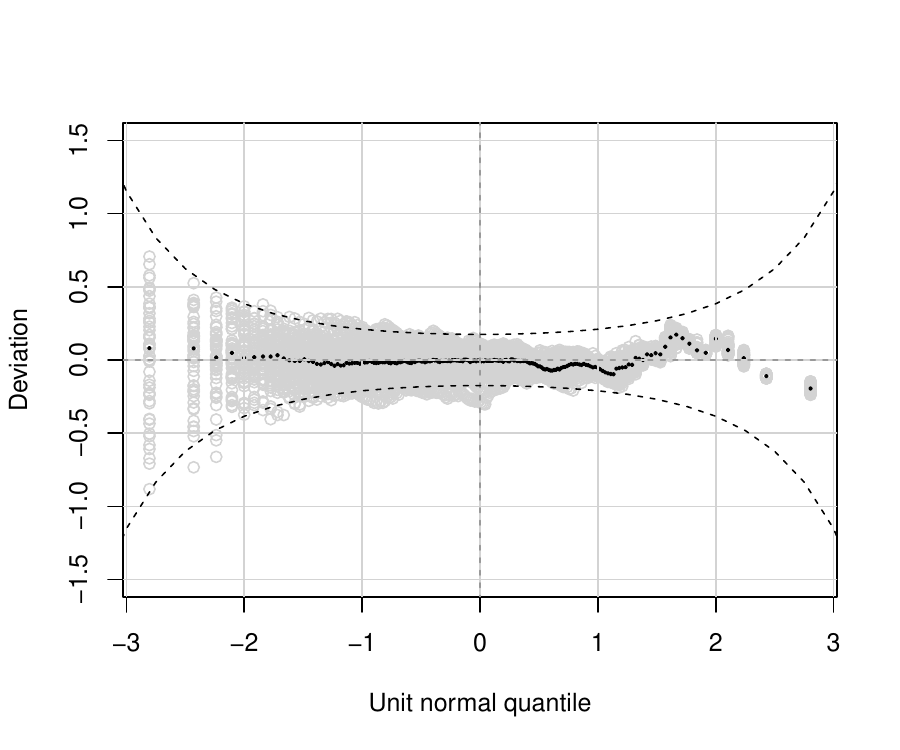}

\caption[Residual diagnostic Worm plots]{Residual diagnostic `Worm' plots for $\mathbbm{1}(D_i = 0)$ (upper panel), $y_{_{I_i}}$ (middle panel) and $y_{_{S_i}}$ (lower panel). The normalized randomized ordered residuals (light gray dots) and their average (black dots) are compared to their (approximate) expected values (horizontal dotted line) and corresponding (approximate) point-wise 95\% confidence intervals (two elliptic curves) [refer to \cite{stasinopoulos2017}, Chapter 12.4, for details]}
\label{fig:rqres}
\end{figure}


\begin{figure}[ht]
    \centering
    \includegraphics[width=0.72\textwidth]
    {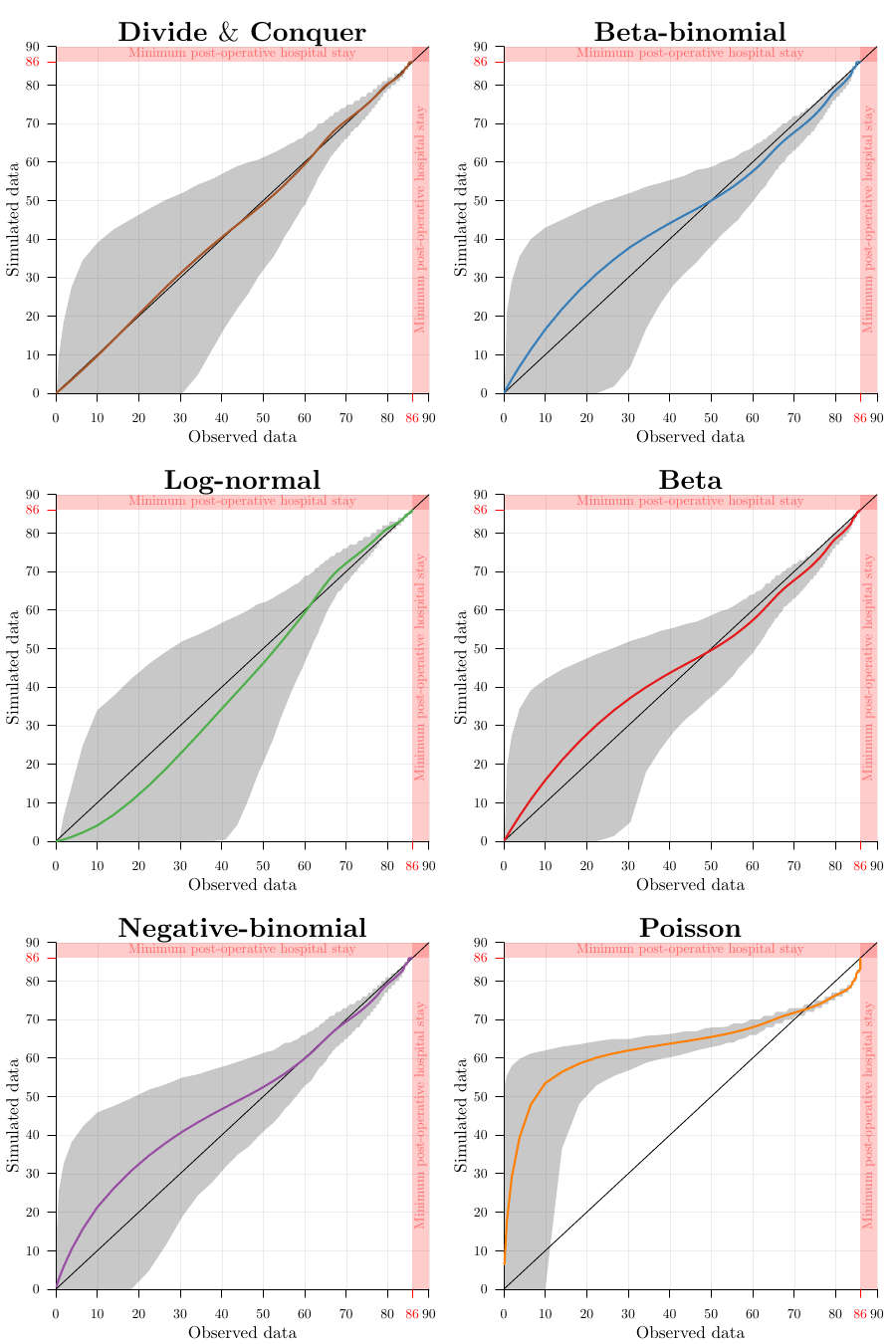}
    \caption{Resample-based quantile--quantile predictive check for the
    fitted DAH model when all methods consider a minimum hospital stay of $\tilde{p}=4$ days: Axes show DAH quantiles in days, evaluated at $250$
  equally spaced probability points in $(0, 1)$. The x-axis gives the
  bootstrap mean of empirical quantiles across $B$ resamples of the
  observed data, and the y-axis gives the bootstrap mean of quantiles
  from data simulated under the fitted model using the resampled
  predictor matrix. The colored line connects these paired points across
  the probability grid; shaded bands give pointwise $95\%$ bootstrap
  intervals ($2.5\%$--$97.5\%$ across replicates) of the model-simulated
  quantiles. The black diagonal line is the identity; alignment of the colored
  line with the identity, within the envelope, indicates that the model
  adequately reproduces the marginal distribution of DAH.}    
  \label{fig:modelcheckB-bootstrap}
\end{figure}


\begin{figure}[ht]
    \centering
    \includegraphics[width=\textwidth]{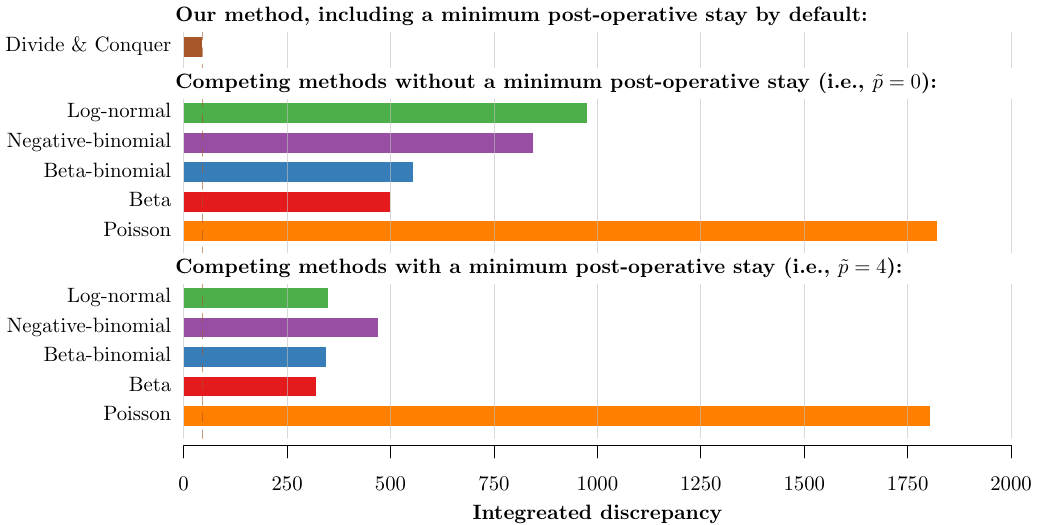}
    \caption{Comparison of the performance of each model: integrated discrepancy between the Q--Q curve and the identity line (x-axis) for each method with and without
minimum postoperative stay (y-axis), where the Q--Q curve corresponds to the
relationship between the bootstrap mean of the empirical quantile from
the observed data and the bootstrap mean of the model-based quantile
from data simulated under the fitted model for each target probability,
as shown by the colored lines in Figures~3 and Web Figure~2. Smaller values of
the discrepancy indicate closer agreement between the model-based and
empirical quantile functions.}
    \label{fig:discrepancy}
\end{figure}


\begin{figure}[ht]
    \centering
    \includegraphics[width=\textwidth]{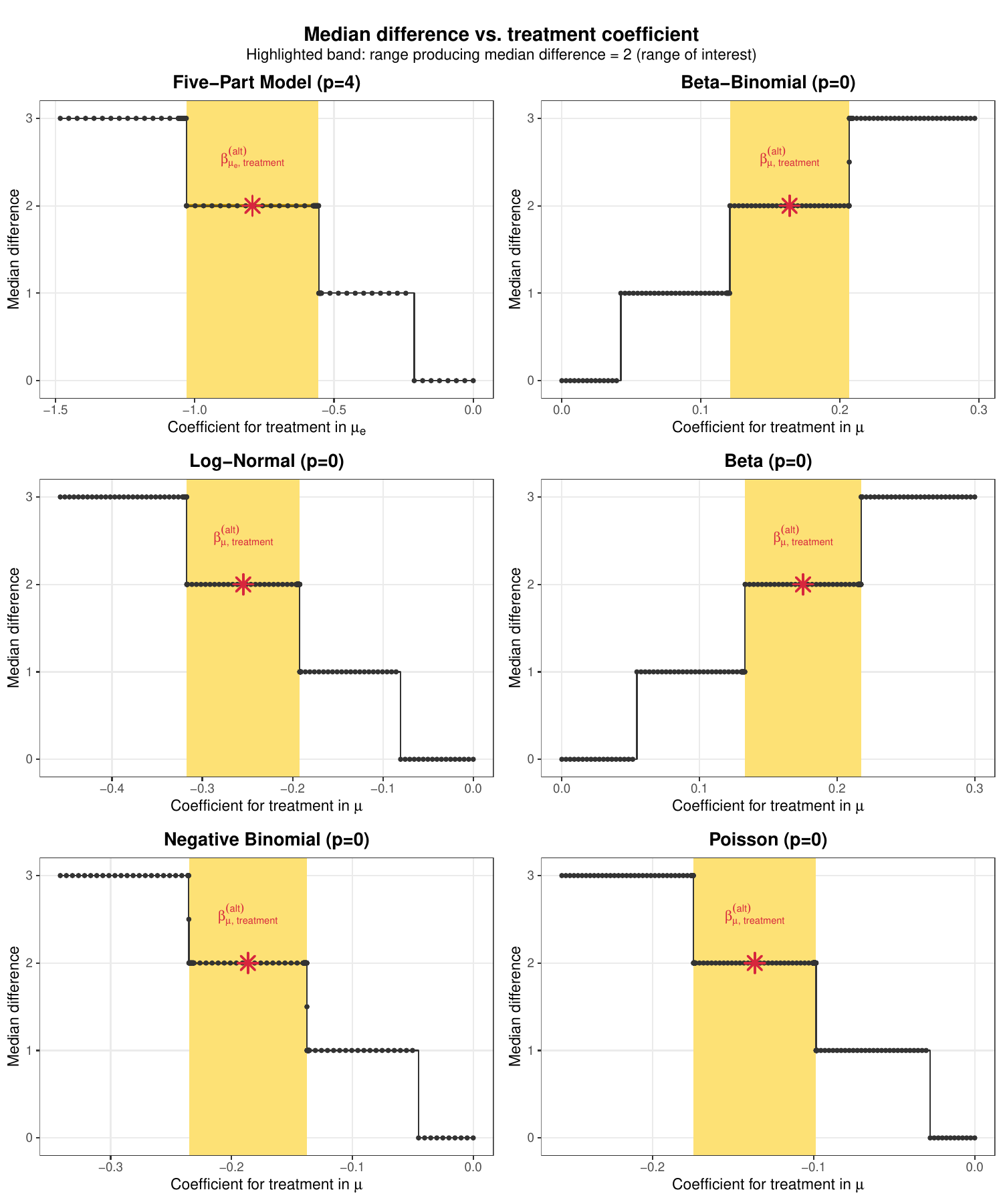}
    \caption{Median difference in DAH between treatment and control arms (y-axis) as a function of the treatment coefficient in the location parameter of each candidate model (x-axis). Each step curve is obtained by evaluating the simulated median difference over a fine grid of coefficient values. The yellow band marks values yielding a median difference of 2, and the red star indicates the midpoint, which is the coefficient value used under the alternative model in the sample size calculation.}
    \label{fig:beta1-stepfunction}
\end{figure}




\label{lastpage}

\end{document}